\RequirePackage{fix-cm}
\documentclass[smallextended]{svjour3}       % onecolumn (second format)
\smartqed  % flush right qed marks, e.g. at end of proof
\usepackage{amsmath,amsfonts,amssymb}
\usepackage[latin1]{inputenc}
\usepackage{graphics} %%
\usepackage{graphicx}
\usepackage{subfigure} 
\usepackage[colorlinks=true, allcolors=blue]{hyperref} 
\usepackage[ngerman,english]{babel} 
\usepackage{lineno}
\usepackage{lscape} 
\usepackage{rotating} 
\usepackage{color}
\usepackage{soul}
\usepackage{xcolor,colortbl}

\usepackage{marvosym}

\journalname{Experimental Astronomy}
\usepackage{tikz,xcolor,hyperref}

\definecolor{lime}{HTML}{A6CE39}
\DeclareRobustCommand{\orcidicon}{%
	\begin{tikzpicture}
	\draw[lime, fill=lime] (0,0) 
	circle [radius=0.16] 
	node[white] {{\fontfamily{qag}\selectfont \tiny ID}};
	\draw[white, fill=white] (-0.0625,0.095) 
	circle [radius=0.007];
	\end{tikzpicture}
	\hspace{-2mm}
}

\foreach \x in {A, ..., Z}{%
	\expandafter\xdef\csname orcid\x\endcsname{\noexpand\href{https://orcid.org/\csname orcidauthor\x\endcsname}{\noexpand\orcidicon}}
}

\begin{document}

%\title{A semi-empirical model for soft proton scattering at grazing incidence from X-ray mirrors}
\title{UVscope and its application aboard the ASTRI-Horn telescope}



\titlerunning{UVscope aboard ASTRI-Horn}        % if too long for running head

\author{Maria~Concetta~Maccarone\textsuperscript{\Email}\orcidA{} \and Giovanni~ La~Rosa\orcidB{} \and Osvaldo~Catalano\orcidC{} \and Salvo~Giarrusso\orcidD{} \and Alberto~Segreto\orcidE{} \and Benedetto~Biondo\orcidF{} \and Pietro~Bruno\orcidG{} \and Carmelo~Gargano\orcidH{} \and Alessandro~Grillo\orcidI{} \and Domenico~Impiombato\orcidL{} \and Francesco~Russo\orcidM{} \and Giuseppe~Sottile\orcidN{} }

\authorrunning{M.C.~Maccarone~et~al.}

\institute{M.C.~Maccarone, G.~La~Rosa, O.~Catalano, S.~Giarrusso, A.~Segreto, B.~Biondo, C.~Gargano, D.~Impiombato, Fr.~Russo, G.~Sottile \at
              IASF-Palermo/INAF, Istituto di Astrofisica Spaziale e Fisica Cosmica di Palermo, Istituto Nazionale di Astrofisica, Via U. La Malfa 153, I-90146 Palermo, Italy \and 
              P.~Bruno, A.~Grillo  \at
              OACT/INAF, Osservatorio Astrofisico di Catania, Istituto Nazionalw di Astrofisica, Via S. Sofia 78, I-95123 Catania, Italy \\\\
              \Email\texttt{\scriptsize{Corresponding author: Cettina.Maccarone@inaf.it}}         
           }
         
\date{Received: date / Accepted: date}

\maketitle

\begin{abstract}
UVscope is an instrument, based on a multi-pixel photon detector, developed to support experimental activities for  high-energy astrophysics and cosmic ray research. The instrument, working in single photon counting mode, is designed to directly measure light flux in the wavelengths range 300-650~nm. The instrument can be used in a wide field of applications where the knowledge of the nocturnal environmental luminosity is required. Currently, one UVscope instrument is allocated onto the external structure of the ASTRI-Horn Cherenkov telescope devoted to the gamma-ray astronomy at very high energies. Being co-aligned with the ASTRI-Horn camera axis, UVscope can measure the diffuse emission of the night sky background simultaneously with the ASTRI-Horn camera, without any interference with the main telescope data taking procedures. UVscope is properly calibrated and  it is used as an independent reference instrument for test and diagnostic of the novel ASTRI-Horn telescope.

\keywords{UVscope -- Night Sky Background -- ASTRI-Horn -- Multianode photomultiplier}

\end{abstract}

\section{Introduction}
\label{introduction}
The atmospheric transparency and the diffuse Night Sky Background (NSB) light are of main interest in the field of high-energy cosmic rays and very-high-energy gamma rays, which, thanks to their interaction with the Earth atmosphere, can be indirectly detected on ground by fluorescence and Cherenkov experiments. For both these fields of research, the atmosphere plays a twofold major role: it is the light emission medium (fluorescence and Cherenkov light production), as well as the transmission medium where the light propagates and attenuates from its point of first interaction to the observation site. Variation in the fluorescence and Cherenkov yield can be induced by variations in the atmosphere's conditions, arising in uncertainty in the cosmic ray energy reconstruction; at the same time, the transmission of the photons from their point of first interaction to the telescope is affected both by scattering and absorption effects, the same effects that influence the NSB light in the UltraViolet (UV) region. The knowledge of the NSB light then provides an important information in characterizing the environment in which the shower takes place and propagates.

The research program of INAF/IASF (Palermo, Italy) devoted to measurements of the NSB started in 1998 with the balloon-born experiment BaBy (Background Bypass)  \cite{Giarrusso 2001} \cite{Giarrusso 2003}, and continued with ground-based observations, firstly with the so-called "Baby on-ground" and later with the UVscope instrument \cite{Catalano 2009}.

UVscope is a light detector working in Single Photon Counting (SPC) mode \cite{Catalano 2008}; its detection unit, calibrated at the INAF Catania Astrophysical Observatory Laboratory for Detectors, COLD 
\footnote{http://cold.oact.inaf.it/cold/index.php/en/}, is a high speed response photomultiplier with quantum efficiency extended to the UV band.
In its first version, the detection unit was a single-anode photomultiplier making use of broad-band filters. Several measurements of diffuse NSB light and of sky transparency were performed at various locations: Madonie Mountains, Italy; Calar Alto Observatory, Spain; ARGO-YBJ, Tibet; Pierre Auger Observatory, Argentina. During spring 2008, UVscope was upgraded in several aspects, first of all adopting as its sensor a multi-anode photomultiplier that, with respect to the previous single-anode unit, exhibits a lower dark current noise and presents moderate imaging properties \cite{Maccarone 2009}. The instrument was completed by a motorized mount and a motorized filter wheel, which allowed to perform programmed measurements at different pointing directions and wavelengths \cite{Maccarone 2011-nima}. Under this configuration, remotely controlled if requested, campaigns of measurements were conducted in Italy (Madonie Mountains) and in Argentina (Pierre Auger Observatory); there UVscope was mounted on the roof of the Los Leones Fluorescence Detector (FD) building \cite{Maccarone 2011-icrc} acquiring data pointing at a fixed altitude-azimuth position in the sky, the same seen by a portion of the "host" FD telescope. Data were acquired contemporaneously but independently from the FD. The comparison between the NSB flux detected by UVscope and that measured by the FD pixels allowed to verify the gain behavior of the FD telescope \cite{Segreto 2011} under real working conditions.

Such successful results and its operational flexibility induced us to consider UVscope as a support instrument for Image Atmospheric Cherenkov Telescopes (IACT) devoted to the very-high-energy gamma-ray astronomy. The occasion has been the ASTRI Project\footnote{ http://www.astri.inaf.it/} \cite{Pareschi 2016} \cite{Scuderi 2018} led by the INAF Institute, whose first goal has been the design, development, deployment and operation of a double-mirror IACT prototype. The telescope, named ASTRI-Horn in honor of the Italian-Jewish astronomer Guido Horn D'Arturo, who pioneered the use of segmented primary mirrors in astronomy \cite{Horn} \cite{Bonoli}, is located on Mt. Etna, Serra La Nave, Italy, at the INAF "M.C. Fracastoro" observing station (37.7\ensuremath{^\circ}N, 15.0\ensuremath{^\circ}E, 1740~m~a.s.l.) equipped with various instrumentation for the monitoring of meteorological and environmental conditions \cite{Maccarone 2013} \cite{Leto 2014}. 

%Fig.\ref{ASTRI-Horn}
\begin{figure}[htb]
\begin{center}
\includegraphics[width=11.5cm]{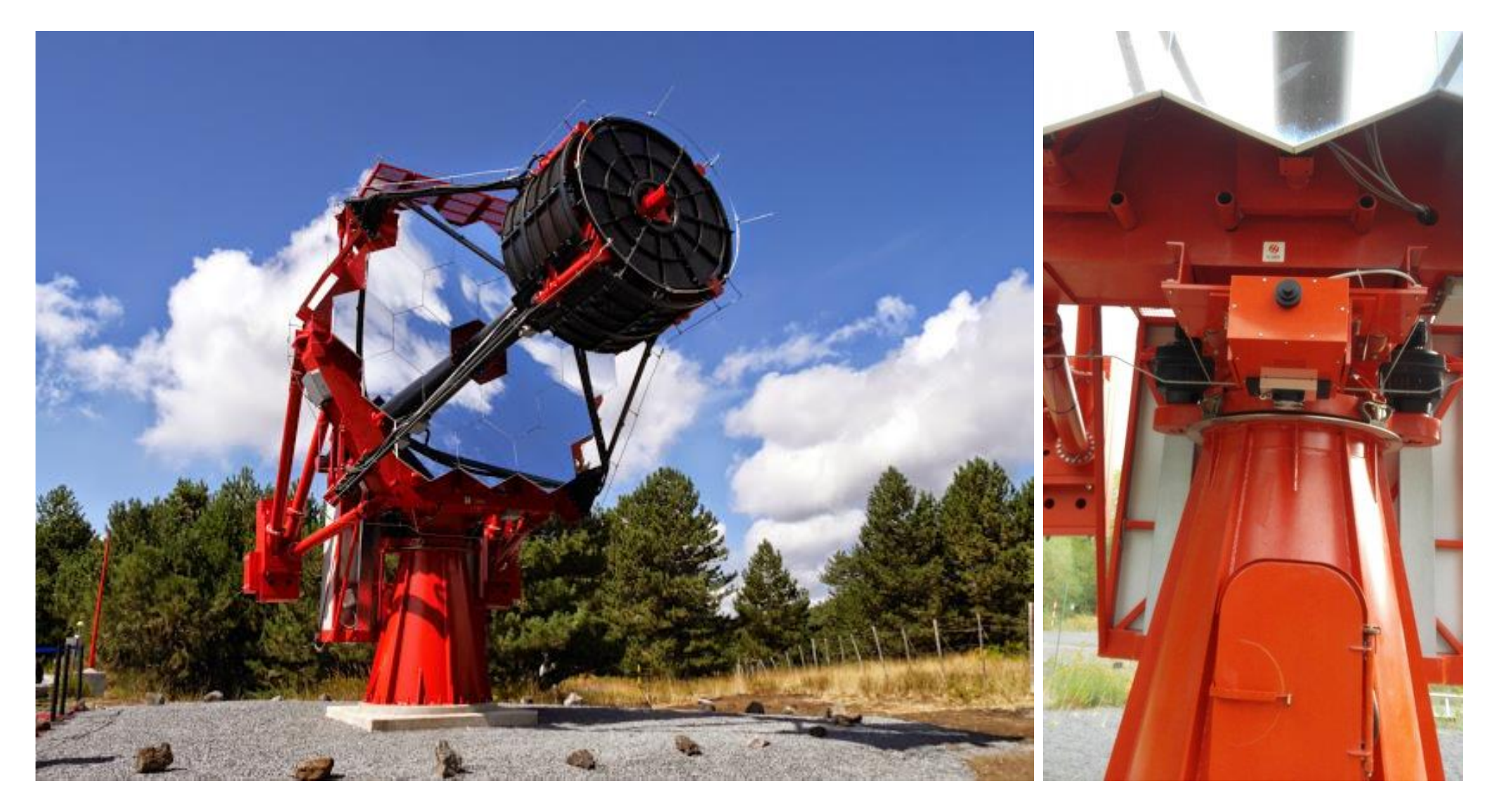}
\caption{ASTRI-Horn, the dual-mirror Cherenkov telescope installed on Mt.Etna, Italy, at the INAF 'M.C. Fracastoro' observing station. On the rigth panel it is visible the small UVscope instrument mounted under the primary mirror structure of the ASTRI-Horn telescope.}
\label{ASTRI-Horn}
\end{center}
\end{figure}

A complete description of the ASTRI Project and ASTRI-Horn telescope can be found in several contributions and references therein (see Lombardi et al. \cite{Lombardi 2020} for a comprehensive bibliography). The ASTRI-Horn telescope (see Fig.\ref{ASTRI-Horn}, left panel) is characterized by an optical system based on a dual-mirror Schwarzschild-Couder design and a camera at the focal plane composed of silicon photomultiplier (SiPM) sensors of 8$\mathrm{\times}$8 pixels, each pixel with a sky-projected angular size of 0.19\ensuremath{^\circ}. The SiPMs, managed by a fast read-out electronics specifically designed, are arranged in 21 detection units, named photon detection modules (PDMs), for a total effective FoV of about 7.6\ensuremath{^\circ} \cite{Catalano 2018}. The ASTRI-Horn telescope is the prototype of nine similar telescopes forming the ASTRI Mini-Array that will be installed at the Teide Astronomical Observatory,  in Tenerife, Canary Islands, Spain \cite{Pareschi 2019}.

The ASTRI-Horn camera front-end electronics is AC coupled and so it blocks any slow varying signal making the telescope blind to the NSB flux (diffuse emission and stars). However, the firmware of the camera continuously performs the analysis of the signal detected by the front-end electronics and periodically provides in output the "variance" of each pixel, which is linearly dependent on the rate of detected photons. The variance technique so allows us to indirectly measure the sky background flux and therefore to investigate variations in the atmospheric attenuation and to monitor the presence of clouds and stars in the telescope field of view. Moreover, as an example, since the positions of the stars are known with a high accuracy, the effective pointing of the telescope can be determined by measuring the shift between the actual positions of the stars and the nominal ones in the images generated by the variance method \cite{Segreto 2019}. 
The front-end electronics of UVscope is DC coupled and so it is able to measure the DC (or slow varying) component of the NSB. The simultaneous measurement of the NSB with UVscope so provides a validation of the variance technique in the ASTRI-Horn camera.

For that primary purpose, an upgraded and simplified version of UVscope has been mounted on the external structure of the ASTRI-Horn telescope.

In this paper, after an overview on the updated UVscope instrument, we detail the data taking and analysis procedures to obtain the diffuse NSB flux. We show diffuse NSB flux profiles evaluated by UVscope  in summer and winter 2018; these profiles confirm the correct behaviour of UVscope and improve our knowledge of the NSB light in the Serra La Nave sky. 

\section{The UVscope instrument aboard ASTRI-Horn}
\label{Uvscope}

The UVscope instrument designed for the ASTRI-Horn is a simplified version of what used in the past, while maintaining the same logic. It basically consists of: a photon detector with its front-end and data acquisition electronics units working in SPC mode, and an interface card for computer connection; a pinhole collimator to regulate the angular aperture of the detector and to protect its sensitive area against stray light; a motorized diaphragm to open/close the entrance pupil during night/day; an air-ventilation system; the power unit. To be operative, the UVscope instrument aboard ASTRI-Horn simply needs a 24~V power and an Ethernet connection.

\begin{figure}[htb]
\begin{center}
\includegraphics[width=11.5cm]{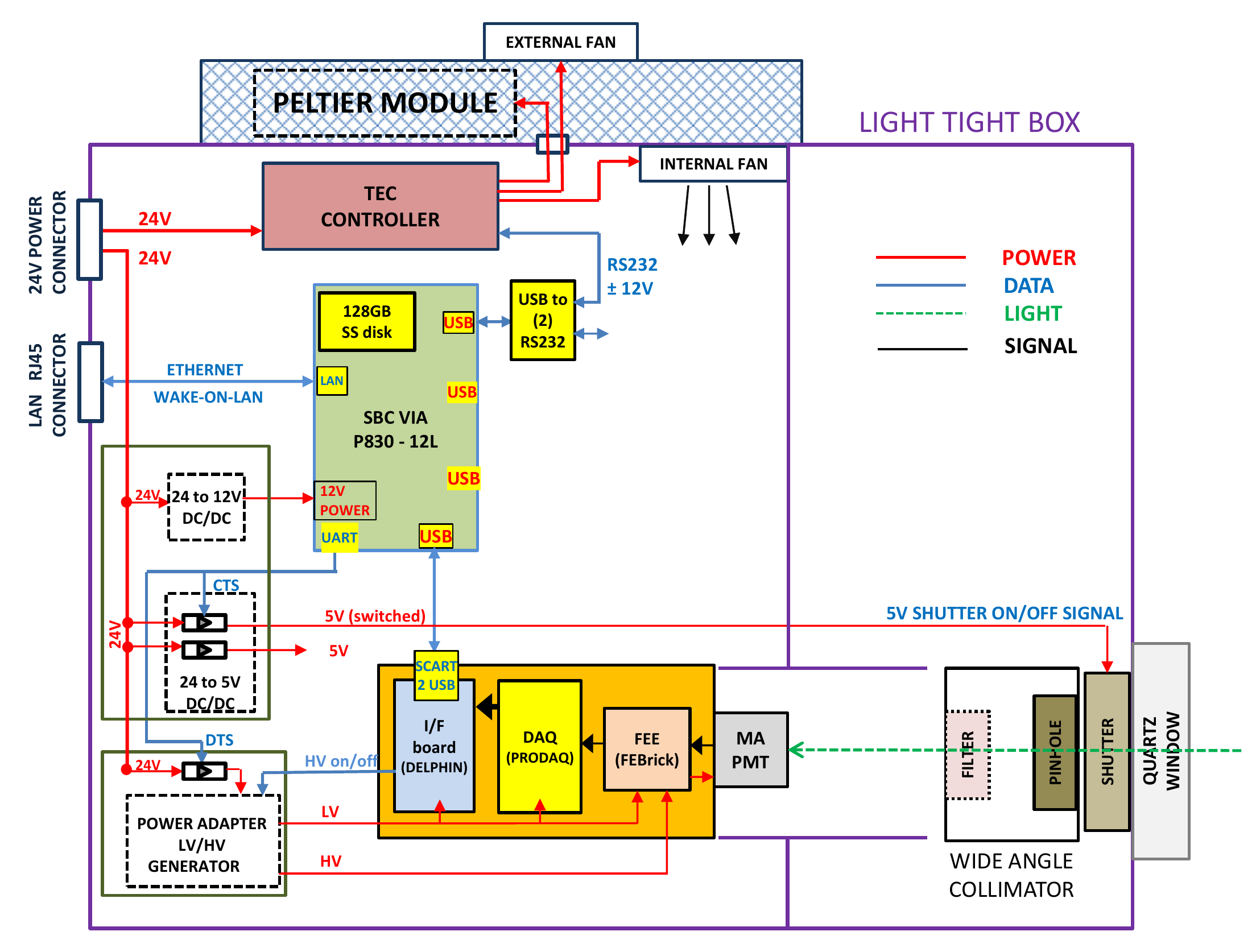}
\caption{The UVscope block-diagram.The lodging for the filter is not used in the current configuration.}
\label{UVscope block diagram}
\end{center}
\end{figure}

The UVscope sensor unit and its electronics have been widely described in a previous paper \cite{Maccarone 2011-nima}; here we briefly report their basic features. The UVscope light sensor is a Multi Anode Photo Multiplier Tube (MAPMT) manufactured by Hamamatsu, series R7600-03-M64
\cite{R7600-M64}, which allows moderate imaging properties with its 64 anodes arranged in a matrix of 8$\mathrm{\times}$8 pixels. The quantum efficiency, extended to the UV band, presents a peak of more than 20\% at 420~nm; the gain is of the order of 3$\mathrm{\times}$10${}^{5}$ at -850~V and the pulse rise time is of 1~ns. The Front-End Electronics (FEE) works in SPC mode \cite{Catalano 2008} that allows to measure the number of output pulses from the photo-sensors corresponding to the number of incident photons with a double pulse time resolution of 10~ns. Signals detected by the FEE are sampled by the programmable Data Acquisition (DAQ) and then processed according to suitable and flexible user algorithms. Fig.\ref{UVscope block diagram} shows the UVscope block diagram.

%Fig.\ref{UVscope block diagram}

\begin{figure}[htb]
\begin{center}
\includegraphics[width=11.5cm]{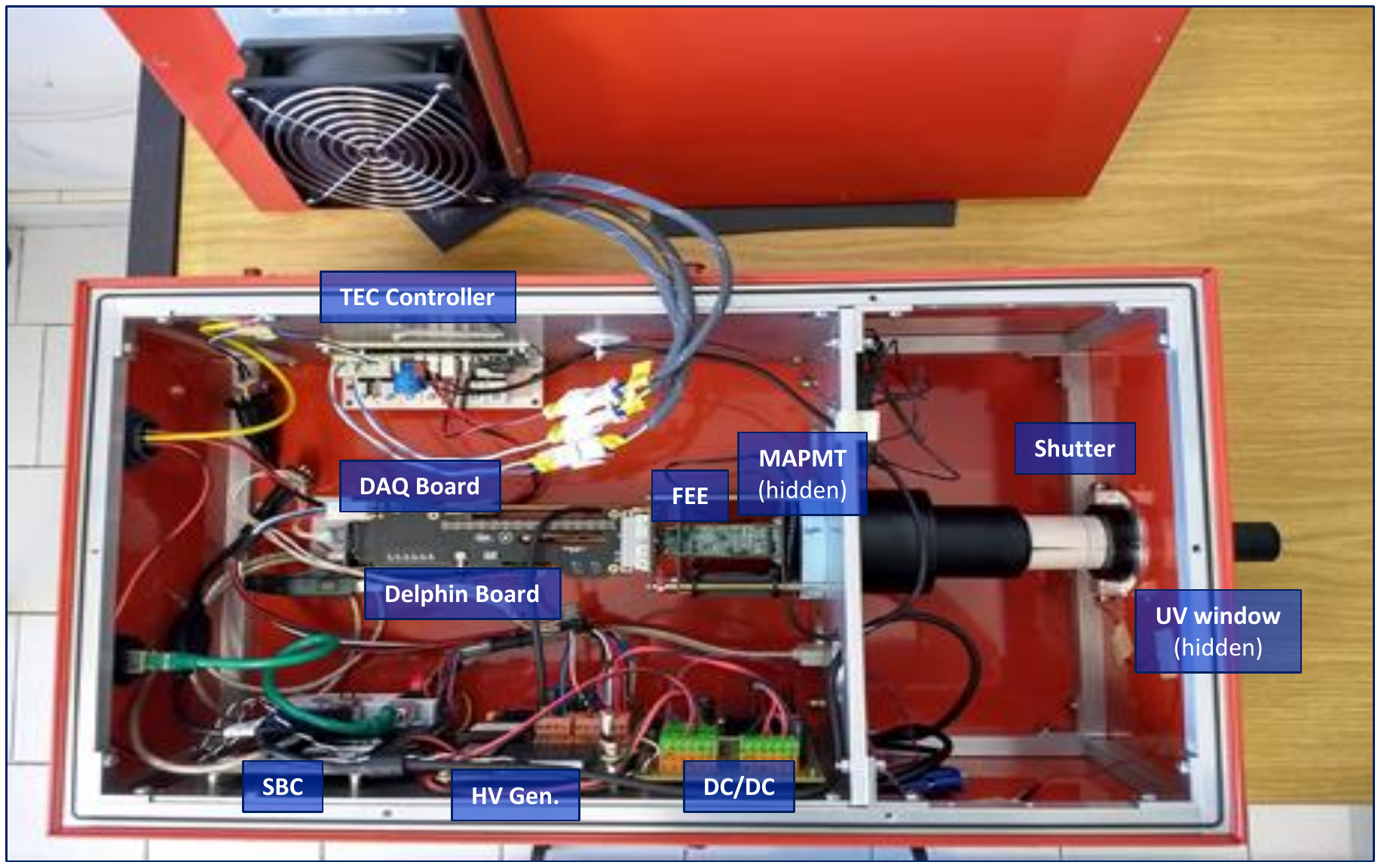}
\caption{Inside UVscope.}
\label{Inside UVscope}
\end{center}
\end{figure}

The UVscope instrument is enclosed in a box (see Fig.\ref{Inside UVscope}), thermal stabilized by an air-to-air cooling unit (TEC) based on Peltier cells that maintains the detection unit at about 20$^\circ$C to strongly reduce (less than 0.7~counts/second per pixel) the dark noise. The air circulation inside the box is guaranteed by the wide inner free zones; moreover, through external outputs the air flux is piped away from the ASTRI-Horn primary mirror. To control the environmental status inside the UVscope box, a set of temperature and humidity sensors are placed in its different zones. 
The box  is equipped with a motorized diaphragm (or shutter) to open/close the entrance pupil during night/day, and with a quartz window, 99\% UV transparent and anti-reflecting coated, externally placed in front of the apparatus' eye.  Eventually, the UVscope apparatus requires a collimator, whose length can vary on the basis of the measurements to be performed and of the desired FoV. A second quartz window is allocated inside the collimator to protect it against unexpected environmental critical conditions during data taking. For a complete protection against undesired shutter failures during data taking and/or environmental critical conditions, a removable plastic cup is applied at the extreme side of the collimator. Fig.\ref{ASTRI-Horn}, right panel, shows UVscope mounted aboard ASTRI-Horn.

%Fig.\ref{Inside UVscope}

The UVscope data read-out is achieved by an internal Single Board Computer (SBC) connected through an interface card, named DELPHIN and entirely developed at the INAF/IASF-Palermo Institute \cite{Russo 2006}, which emulates a disk unit. The SBC (Windows7 Operating System) is devoted to the initialization of all UVscope subsystems and to the activation of the Local Area Network (LAN) connection. In such a way, any computer remotely connected to UVscope through the LAN network can perform all the necessary checks and operations (temperature and humidity control, Peltier status,  switching on/off the UVscope electronics, open/close the shutter, start/stop the data acquisition,{\dots}) and eventually download all the acquired data for the off-line analysis.

The data acquired with UVscope are stored in ASCII files. The header maintains general information as date and start time of the acquisition, electronics setup, integration time window. Then the acquired data are registered, pixel by pixel, as number of counts in the pre-selected integration time, accompanied by the value of the temperature of the electronic boards inside the instrument box and the high voltage detected between anode and last dynode.

Although originally designed to operate as a standalone instrument, in the ASTRI-Horn system UVscope is viewed as an auxiliary device integrated in the ASTRI-Horn control software, named MASS  (Mini Array Software System) \cite{Tanci 2016}.  MASS is based on the ACS (Alma Common Software) framework and on the industrial standard OPC-UA (Open Platform Communications - Unified Architecture) communication protocol. An OPC-UA server has been realized and installed on the UVscope SBC, where  it runs as a Windows Service. The server exposes the UVscope data to the MASS hight level software and, in particular, it implements  two main commands:  UVM\_SBC\_SHUTDOWN and UVM\_SBC\_RESTART in order to execute the shutdown and restart of the Windows operating system running on SBC. Moreover, an ACS component, that includes an OPC-UA client, has been realized in order to obtain the interaction between UVscope and the main ASTRI-Horn telescope Graphic User Interface (GUI) and to monitor the UVscope data. A suitable ACS component, named monitorDB, eventually performs the storage of all UVscope data (configuration parameters, housekeeping and scientific data) on the ASTRI-Horn Telescope Monitoring and Configuration Data Base, TMCDB.

A set of home-made software programs and utilities completes the system, both for the data acquisition and quick-look and for the data analysis. 

\section{Preparing UVscope for the data acquisition}
\label{configuration}

\subsection{The sensor unit and its global efficiency}
\label{efficiency}

The efficiency of the MAPMT detection unit used in our acquisitions, namely PM0091, was previously measured in the laboratory \cite{Maccarone 2011-nima} by comparison with a photodiode calibrated and certified by the National Institute of Standards and Technology (NIST)\cite{NIST}. After a short period of usage (less than 30 hours) and before to be the mounted on UVscope aboard ASTRI-Horn, the efficiency of PM0091 was again measured, showing no aging. Once again, the PM0091 global average efficiency (comprehensive of collecting, trigger, quantum efficiencies and quartz windows) resulted of 11.61\% as reported in Table~\ref{tab:MAPMT features} together with the PM0091 features and operational values applied on-field at Serra La Nave. The value of some parameters are commanded through the GUI at the beginning of each acquisition. As an example, the integration time is provided in units of 10~ms, the minimum integration time window that UVscope can perform.

\begin{table}[ht]
\caption{\label{tab:MAPMT features}The MAPMT operational parameters values for UVscope mounted aboard ASTRI-Horn.}
\begin{tabular}{|p{6cm}|p{1.6in}|} \hline
\textbf{Component or Parameter} & \textbf{Code or Value} \\ \hline
UVscope sensor unit, MAPMT  mod. R7600-03-M64 & PM0091 \\ \hline
MAPMT layout  & 8 pixel x 8 pixel  \\ \hline
Integration Time Window & 1000 ms  (as 100 x 10 ms unit) \\ \hline
HV cathode & -850 V \\ \hline
Threshold & 6 mV \\ \hline
Average dark noise & 0.7 counts/second per pixel \\ \hline
Gain linearity & at least till 50 MHz \\ \hline
Global average efficiency: \newline sensor unit with two quartz windows & ... \newline 11.61 \% (calibrated in lab)\\ \hline
\end{tabular}
\end{table}

\subsection{The gain uniformity map}
\label{gain}
The PM0091 gain uniformity, pixel by pixel, has been derived from acquisitions "flat-field". By putting a fluorescent paper in the inner part of the collimator cup and opening the shutter, the PM0091 sensor was uniformly illuminated for an adequate time (few minutes are sufficient to reach a plateau). The counts per second of each UVscope pixel have then been normalized to the content of a pixel chosen far from the border of the sensor and presenting the most constant behaviour along time. This procedure has been repeated before several observation nights and the results indicate that the gain uniformity is maintained constant within 4\% along all the period under examination (about 10 months). The gain uniformity map has been used in the data analysis process.

\begin{figure}[htb]
\begin{center}
\includegraphics[width=11cm]{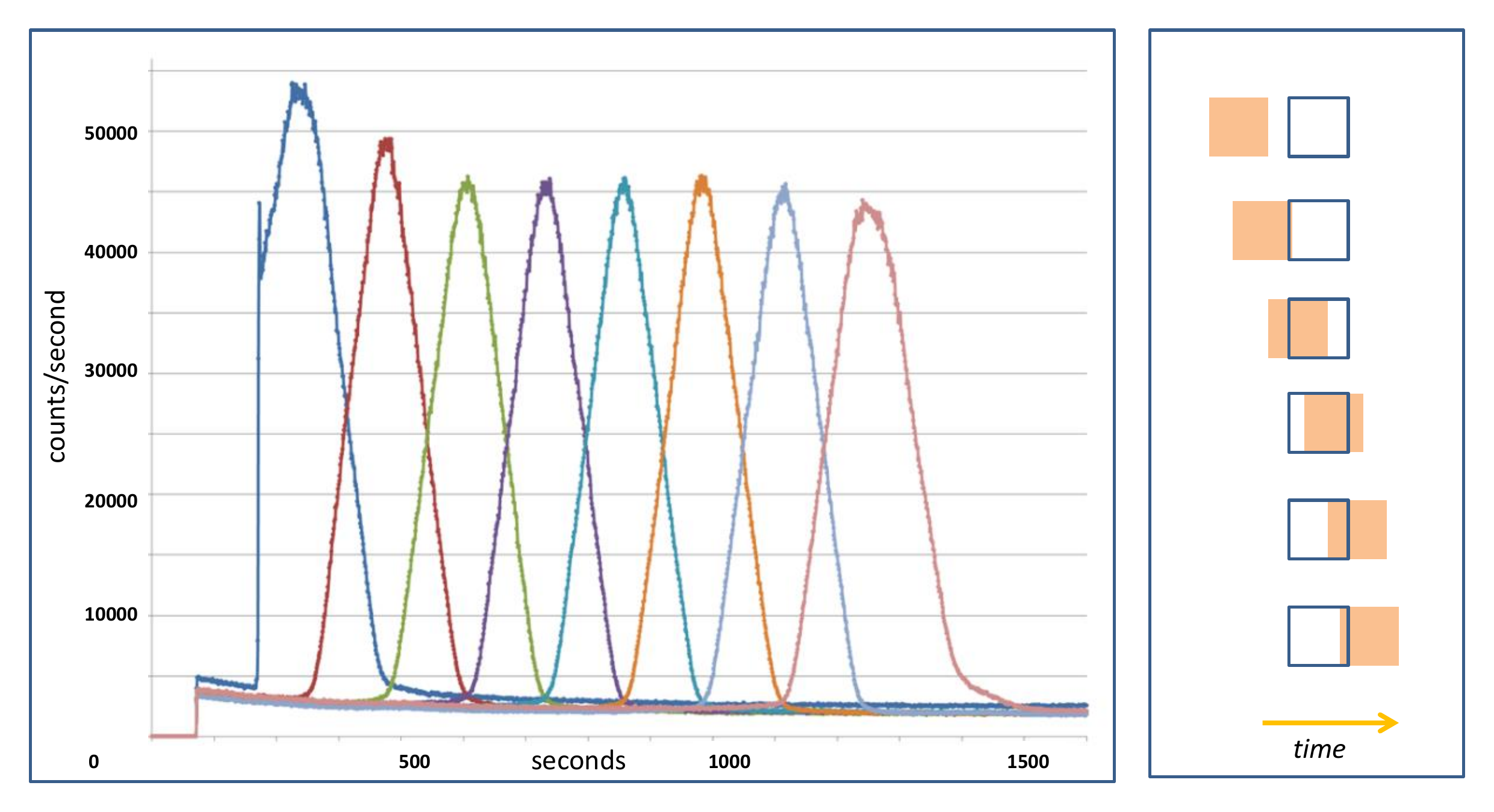}
\caption{Time profile of count-rate during the passage of Jupiter in the UVscope FoV. Left panel: count-rate for each pixel travelled by the Jupiter pencil beam. The entire travel path interested 8 UVscope pixels along a horizontal line in about 1200 seconds. Right panel: schematic representation of the Jupiter pencil beam passage across a single pixel.}
\label{Jupiter passage}
\end{center}
\end{figure}

Nevertheless, despite the gain equalization, it should be emphasized that an intrinsic difference in efficiency of the order of 50\% is always present between the internal pixels and those along the external perimeter of the PM0091 sensor. This is due to the extension of the photocathode, such that the photoelectrons emitted on the edge are focused on the outermost pixels. As an example, Fig.\ref{Jupiter passage} shows the count-rate for each of the pixels travelled by the Jupiter pencil beam, along its passage through the UVscope FOV during a test campaign, May 2017. As the input pin-hole is square, as well as the MAPMT pixels, and of the same dimensions, then the light curve, for each MAPMT pixel, should be a perfect triangle. The first peak, corresponding to the external ring, shows a higher count rate, as expected. The peak of the second pixel is higher than the others only because the PM0091 efficiency is greater in that area, while its physical dimensions (look at the width of the triangle) are exactly the same as those of the central pixels. The most interesting curve is that of the last pixel where it is clear that the triangle is wider (therefore greater physical geometric dimensions) even if the peak is slightly lower (therefore the efficiency of the PM0091 is slightly lower). This pixel has then the particularity of having a higher NSB count rate (the area of the triangle) even if its efficiency (the peak) is lower. These effects will be taken into account depending on the data analysis purpose; in the evaluation of the diffuse NSB light, the external pixels will be excluded, as described in the following (section \ref{useful mask}).

%Fig.\ref{Jupiter passage}

\subsection{The dark counts.}
\label{dark noise}
\begin{figure}[htb]
\begin{center}
\includegraphics[width=11cm]{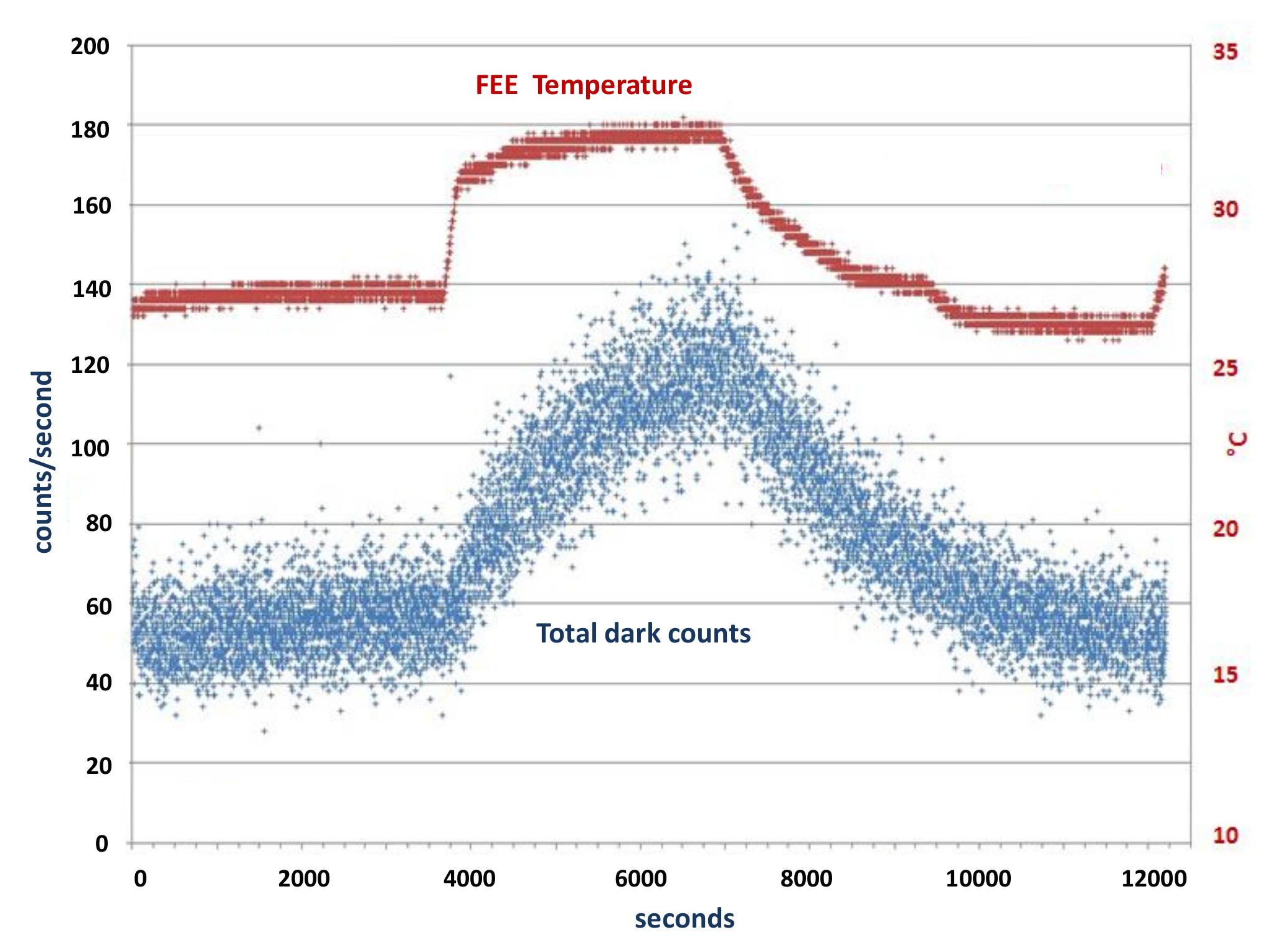}
\caption{Time profile of the FEE temperature (top curve) and of the total dark counts (bottom curve) on all 64 pixels as checked in lab (see text for details).}
\label{dark counts in lab}
\end{center}
\end{figure}

Dark counts depend on temperature but, since UVscope is thermally stabilized by the Peltier system, such noise is negligible. This behaviour was firstly checked in lab (see Fig.\ref{dark counts in lab}): maintaining the FEE temperature around 27\ensuremath{^\circ}C, the total dark noise (for all 64~pixels) resulted of about 50 counts per second, as expected on the basis of previous PM0091 calibration measurements.

%Fig.\ref{dark counts in lab}

Such behavior has then been verified on-field analyzing short dark acquisitions (shutter closed) taken just before the scientific runs; eventually, a conservative value of 0.7 dark counts per pixel per second has been used in the analysis procedure. As an example, we report the values obtained in June 2018. During the dark run (shutter closed, 255 acquisitions), the FEE temperature maintained an average value of 23.54\ensuremath{^\circ}C with a standard deviation of 0.23\ensuremath{^\circ}C. The average dark counts per pixel was of the order 0.7~counts per second. During the subsequent scientific run, the temperature of the FEE decreased to around 19\ensuremath{^\circ}C (St. Dev. 0.17\ensuremath{^\circ}C) consequently decreasing the average dark counts per pixel. The use of an average dark value of 0.7~counts/second per pixel is therefore conservative, despite being 'negligible' for the purposes of the analysis. Values of the same order were found along the entire period here analyzed.

\subsection{The geometrical parameters}
\label{geometry}
\begin{figure}[htb]
\begin{center}
\includegraphics[width=11cm]{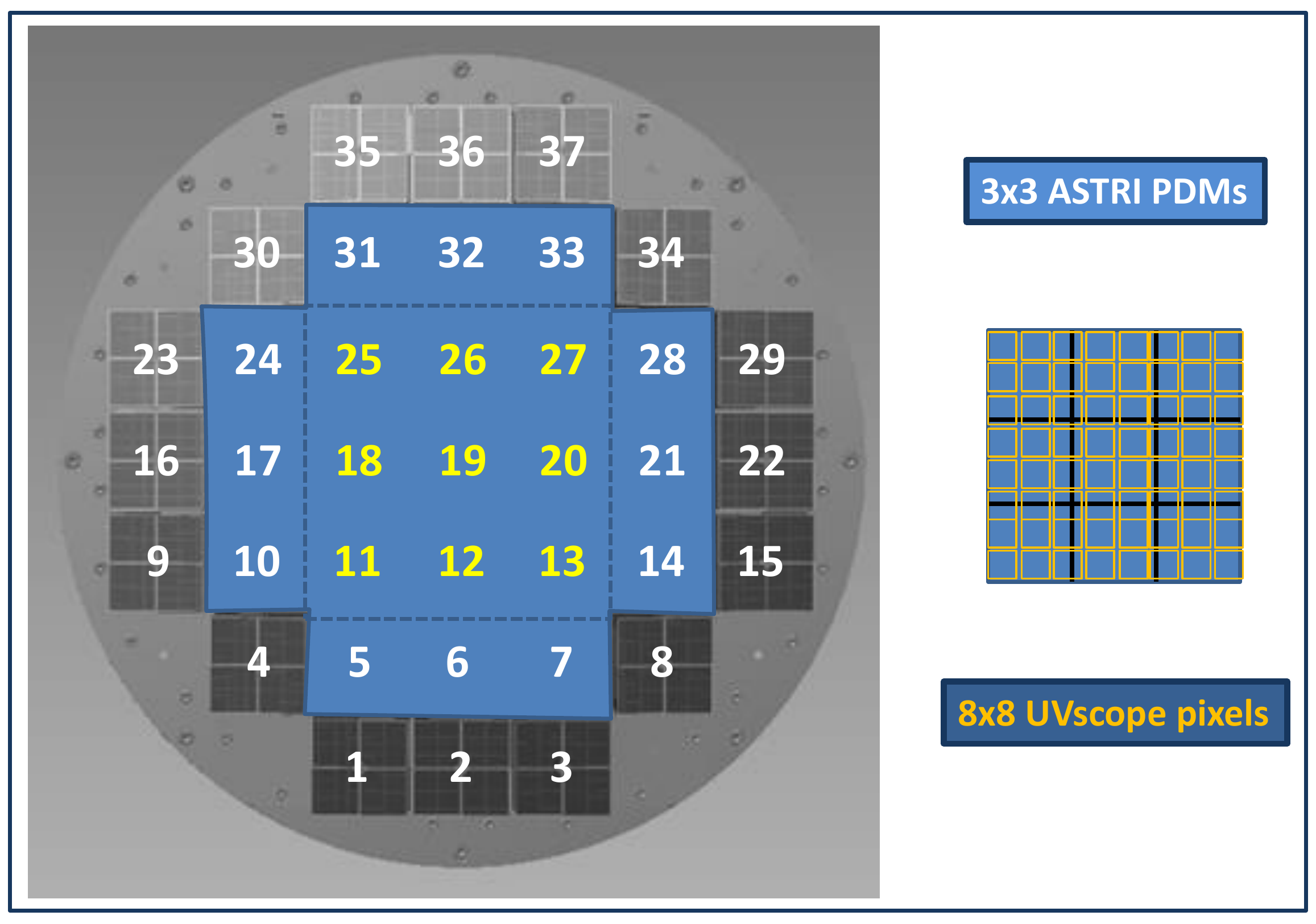}
\caption{The ASTRI-Horn focal surface and UVscope dimensional correspondence, with respect to their FoV. The numbers on the left panel code the ASTRI-Horn camera PDMs while the dashed square around the PDMs in yellow roughly identify the FoV common to UVscope whose pixels are schematized in the right panel.}
\label{FoV UVscope and ASTRI}
\end{center}
\end{figure}

UVscope acquires data simultaneously with the ASTRI-Horn camera, pointing the same sky region independently from the main telescope.  To relate the UVscope measurements as much as possible with the central area of the ASTRI-Horn camera, a collimator equipped with a square entrance pupil has been used; the  distance between pupil and photocathode has been settled at 238.1~mm so to achieve an angular aperture of UVscope pixel of 0.55\ensuremath{^\circ}. Under these geometrical conditions, summarized in 
Table~\ref{tab:UVscope geometrical parameters}, the UVscope field of view covers the internal area of 3$\mathrm{\times}$3 PDMs of the ASTRI-Horn camera, being the aperture of each UVscope pixel equal to about 3$\mathrm{\times}$3 pixels of the ASTRI-Horn camera (0.19\ensuremath{^\circ} angular pixel size), as schematized in Fig.\ref{FoV UVscope and ASTRI}.

\noindent
\begin{table}[ht]
\caption{\label{tab:UVscope geometrical parameters}The MAPMT operational parameters values.}
\begin{tabular}{|p{0.8in}|p{1.5in}|p{0.7in}|p{0.75in}|} \hline
\textbf{Parameter} & \textbf{Meaning} & \textbf{Value} & \textbf{Units}  \\ \hline
\textbf{L${}_{pixel}$} & Effective pixel linear size  \textbf{} & 2.3 $\mathrm{\times}$ 2.3 & mm $\mathrm{\times}$ mm  \\ \hline
\textbf{A${}_{pixel}$} & Effective pixel area  \textbf{} & 5.29 & mm${}^{2}$  \\ \hline
\textbf{L${}_{pupil}$} & Pupil linear size  \textbf{} & 2.0 $\mathrm{\times}$ 2.0 & mm $\mathrm{\times}$ mm  \\ \hline
\textbf{A${}_{pupil}$} & Pupil area  \textbf{} & 4.0 & mm${}^{2}$  \\ \hline
\textbf{L${}_{PMT}$} & Effective PMT linear size  \textbf{} & 18.1 $\mathrm{\times}$ 18.1 & mm $\mathrm{\times}$ mm  \\ \hline
\textbf{D${}_{pupil}$} & Distance between pupil and photocathode  \textbf{} & 238.1 & mm \\ \hline
\textbf{FFoV} & Full Field of View  \textbf{} & 4.35 & deg \\ \hline
\textbf{S${}_{pixel}$} & Angular pixel aperture  \textbf{} & 0.55 & deg \\ \hline
\textbf{Platescale} & Ratio between angular and linear pixel size  \textbf{} & 0.2403 & deg / mm \\ \hline
\textbf{$\Omega_{pixel}$} & Pixel solid angle  \textbf{} & 9.3312e-5 & sr  \\
\textbf{} &  & 0.3063 & deg${}^{2}$  \\ \hline
\textbf{GF${}_{pixel}$} & Geometrical Factor  \textbf{} & 3.73248e-4 & mm${}^{2}$ $\mathrm{\times}$ sr \\
\textbf{} &  & 1.22529 & mm${}^{2}$ $\mathrm{\times}$ deg${}^{2}$  \\ \hline
\end{tabular}
\end{table}

%Fig.\ref{FoV UVscope and ASTRI}

\subsection{Alignment between UVscope and ASTRI-Horn camera.}
\label{alignment}

\begin{figure}[htb]
\begin{center}
\includegraphics[width=11cm]{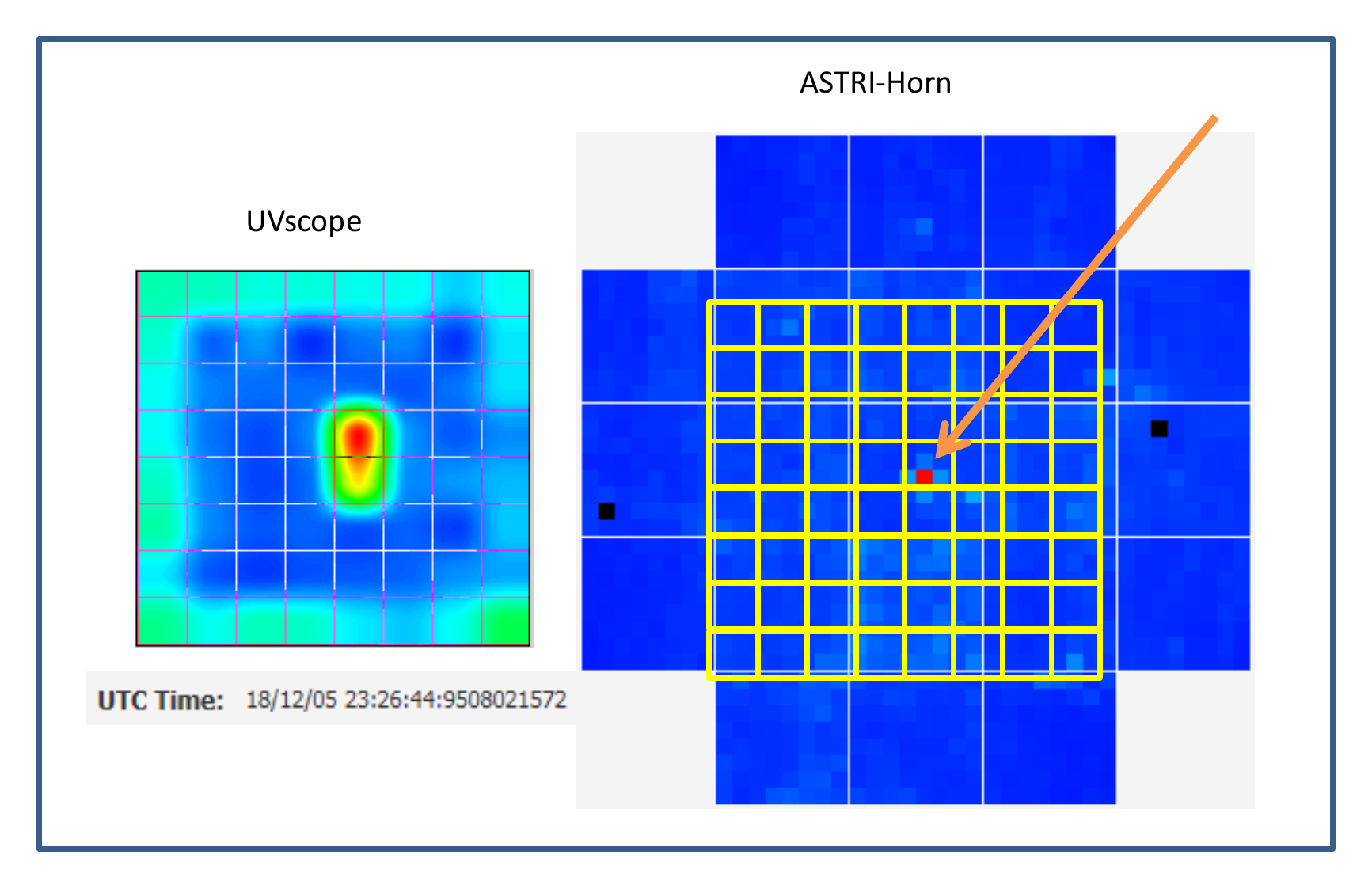}
\caption{Snapshot of a Zeta Tauri tracking observation as acquired by UVscope (left) and by ASTRI-Horn camera (right) in variance mode (see text for details).}
\label{alignment tracking}
\end{center}
\end{figure}

The angular aperture of the UVscope pixel (0.55\ensuremath{^\circ}) with respect to the ASTRI-Horn camera pixel (0.19\ensuremath{^\circ}) 
requires an alignment better than a quarter of the ASTRI-Horn camera pixel. Fig.\ref{alignment tracking} shows a snapshot of the observation of Zeta Tauri. The left panel shows the UVscope data map (smoothed version for an easier visualization). On the right panel there is the map of the variance data acquired by ASTRI-Horn;  superimposed on it there is the UVscope FoV expressed as a grid of its MAPMT pixels. The arrow indicates the pixel where the star is imaged  in the ASTRI-Horn variance map (PDM~\#19, pixel~\#37). In the UVscope data map, the star is mainly detected near the center (pixel~\#37) with some skidding toward pixel~\#29). This position relationship has been confirmed in all stellar tracking acquisitions.

%Fig.\ref{alignment tracking}

\begin{figure}[htb]
\begin{center}
\includegraphics[width=11cm]{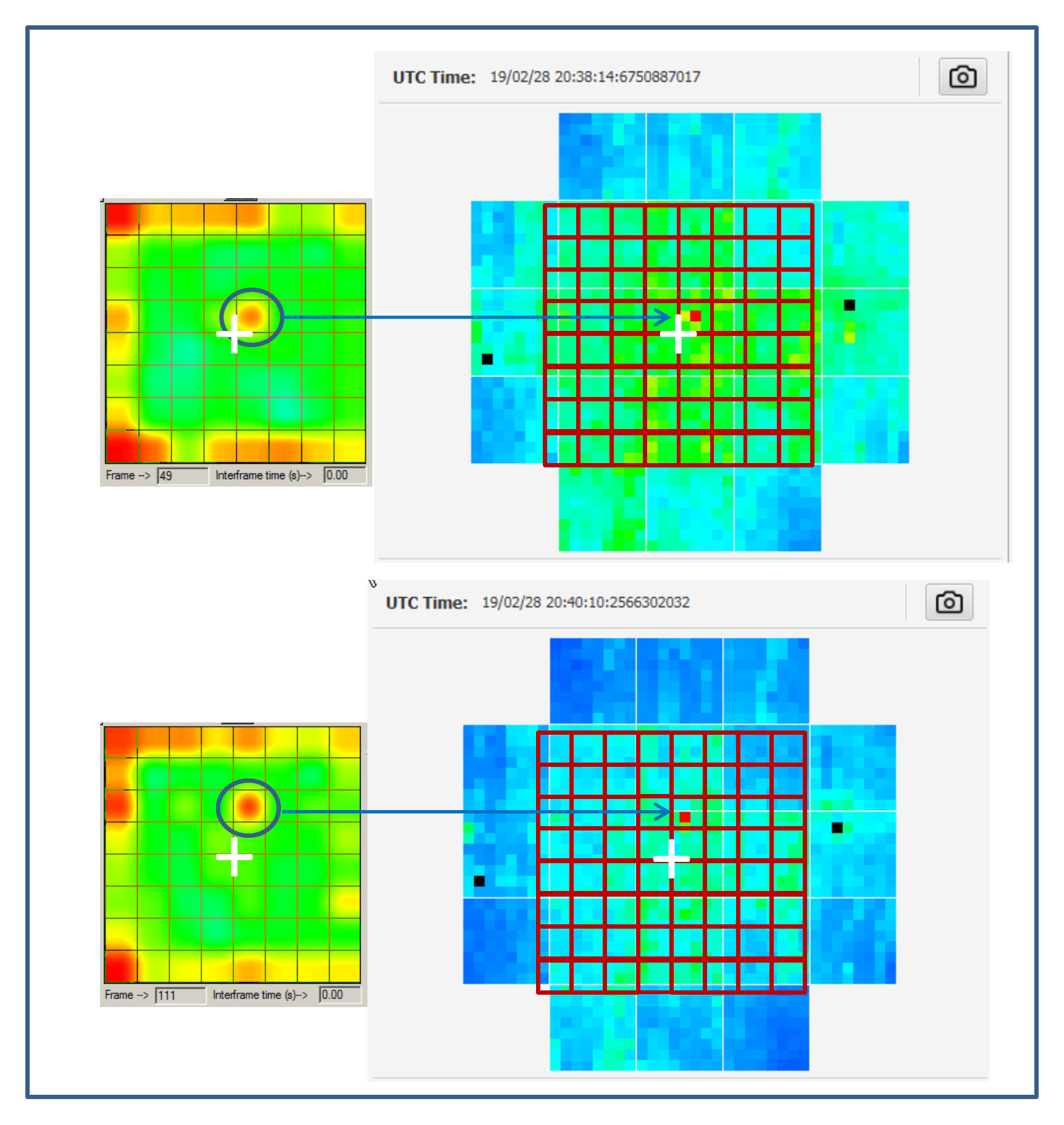}
\caption{Staring observation near Beta Leonis Minoris. Left: UVscope data map. Right: ASTRI-Horn variance map (see text for details).}
\label{alignment staring}
\end{center}
\end{figure}

This tracking check provides information only on one pixel, the one where the bright star is located. To have a notion, albeit approximate, of the alignment in question, it has been necessary to use fixed Altitude-Azimuth coordinate observations (staring) near the star so to be able to view its path in the two systems. Fig.\ref{alignment staring} shows what happens when the telescope is pointed in staring near the Beta Leonis Minoris star, after a short tracking of the star itself. The movement on the planes of the two detectors occurs vertically thus demonstrating that the UVscope grid is positioned to the required alignment as superimposed on the focal plane of the ASTRI-Horn camera (white cross: center of UVscope).

%Fig.\ref{alignment staring}

\section{The data analysis procedure for the diffuse NSB flux evaluation}
\label{data analysis}

The NSB light depends on the latitude and longitude of the observing point and on the atmosphere conditions, including humidity, pressure, dusts, clouds, airglow, moonlight, stars and planets, and so on \cite{Leinert 1998}. We note that UVscope and ASTRI-Horn are installed on the slope of the Etna Volcano; the local atmosphere is often influenced by eruptive dust and clouds of smoke that depend on the wind and whose effect depends on the pointing of the telescope. Such events related to eruptive conditions therefore contribute in decreasing the transparency of the night sky so decreasing its apparent magnitude; as consequence, they increase the diffuse sky brightness and therefore increase the night sky background level.

In conditions of moonless clear nights (lack of clouds), the diffuse NSB value is defined as the brightness of the sky, excluding the brightest stars, when the Sun is well below the horizon (\ensuremath{<}-18\ensuremath{^\circ},  astronomical twilight). To exclude the brightest stars from the UVscope field of view the concept of an "useful mask" is used in the analysis. 

\subsection{The useful mask}
\label{useful mask}

For the diffuse NSB evaluation, the 28 pixels of the external frame, whose contents are influenced by effects on the edges, are excluded from the analysis. Moreover, the image of a bright star  appears in UVscope in maximum 4 pixels (if the star is located across more pixels); such pixels are also excluded from the analysis. The remaining 32 pixels will form the  "useful mask".

%Fig.\ref{useful mask Eta Herculis}
\begin{figure}[htb]
\begin{center}
\includegraphics[width=11cm]{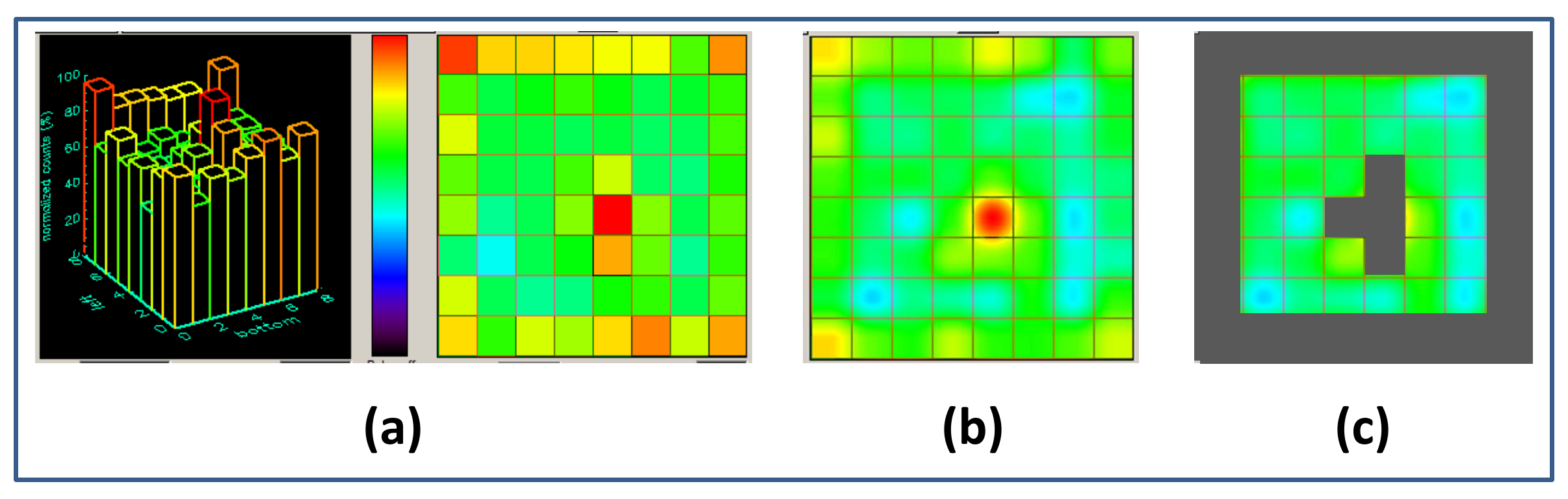}
\caption{Definition of the UVscope useful mask for the evaluation of the diffuse NSB light around Eta Herculis. (a) Counts per second per pixel in 3-D and 2-D representation. (b) Smoothed  version of the counts map. (c) Useful mask: the pixels in gray are NOT part of the useful mask.}
\label{useful mask Eta Herculis}
\end{center}
\end{figure}

%Fig.\ref{useful mask Vega}
\begin{figure}[htb]
\begin{center}
\includegraphics[width=11cm]{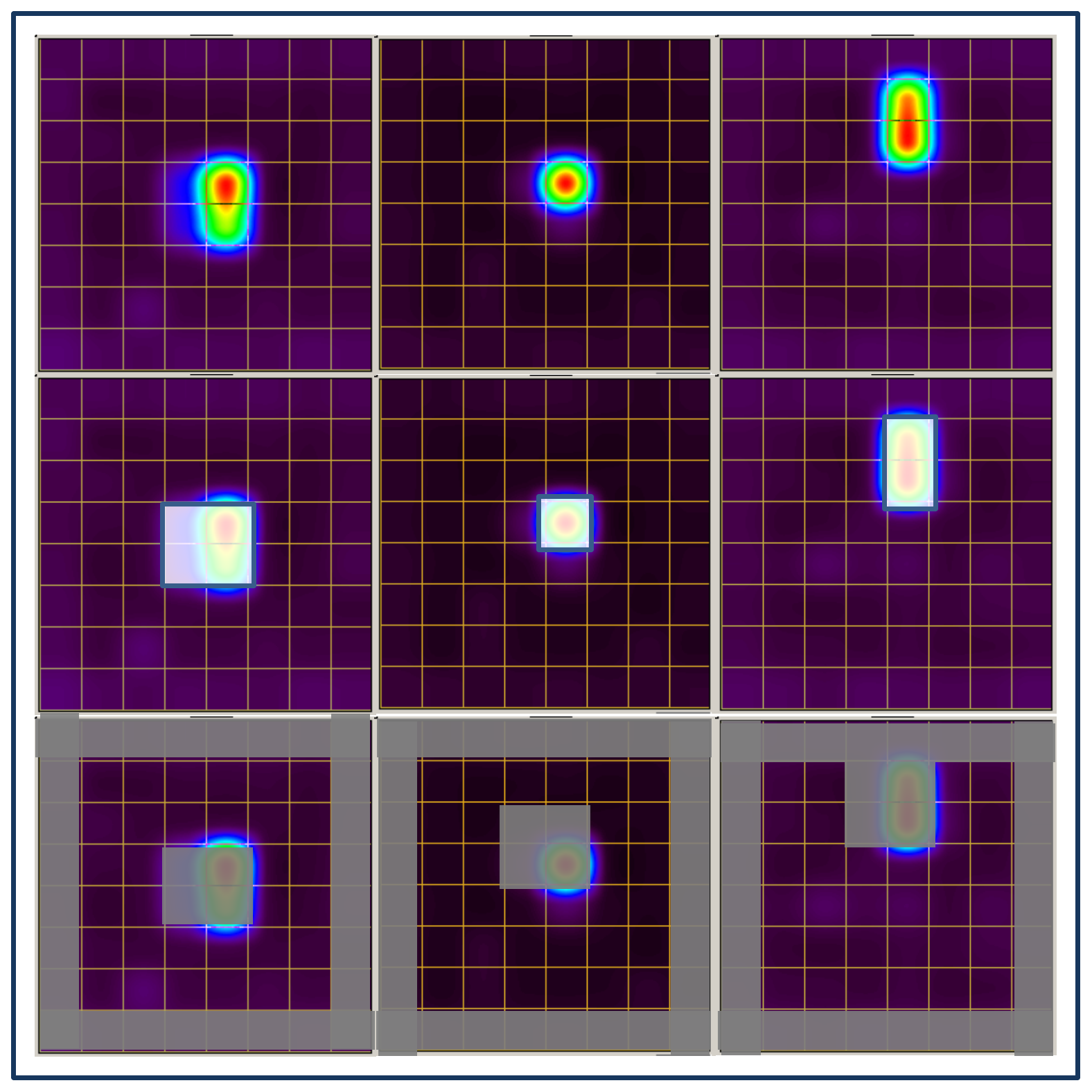}
\caption{Dynamic definition of the useful mask during the passage of Vega across the UVscope FoV. Top panel: smoothed counts map per pixel per second at $t1<t2<t3$ time. Middle panel: the white boxes indicate the pixels where Vega is present. Bottom panel: the dynamic useful mask at $t1, t2, t3$ time; the pixels in gray are NOT part of the useful mask.}
\label{useful mask Vega}
\end{center}
\end{figure}

The useful mask is automatically determined in the analysis, acquisition by acquisition, excluding both the 28 pixels of the external frame and the first 4 pixels with highest content as detected in the 6\ensuremath{\times}6 pixels central zone. Some examples are given in Fig.\ref{useful mask Eta Herculis} and Fig.\ref{useful mask Vega}, where the smoothed version of the image is given only for display purposes. In both the figures, the pixels in gray represent the excluded ones for the given acquisition.

\subsection{The diffuse NSB evaluation procedure}
\label{nsb procedure}

The data acquired by UVscope are registered, pixel by pixel, as number of counts in a pre-selected integration time. In our case we have set a rate of one acquisition each second. To evaluate the mean flux of the diffuse NSB, \textit{$<NSB>(t)$}, the UVscope data, pixel by pixel, are cleaned of dark counts (albeit very low), normalized to the gain uniformity map of its sensor unit, scaled for the global average efficiency of the sensor itself and, after selection in the useful mask, used for the evaluation of:\\

\noindent
\[<NSB>\left(t\right)=\frac{{CTS}^*(t)\cdot {10}^{-9}}{{GF}_{pixel}\cdot {10}^{-6}\cdot <{\varepsilon }_{total}>}\ \ \ \ \ \ \ \ \ \ \left[\frac{photons}{m^2\cdot ns\cdot sr}\right]\]
where:
\[{CTS}^*\left(t\right)=\ \frac{1}{Npix}\ \sum^{Npix}_{k=1}{\frac{CTS\left(k,t\right)-<dark(T\left(t\right))>}{equ(k)}}\ \ \ \ \left[\frac{counts}{second}\ per\ pixel\right]\]

\noindent where \textit{k} indicates each of the \textit{N${}_{pix}$ }forming the useful mask, \textit{CTS${}_{k}$} are the counts in the pixel \textit{k} at time \textit{t}, \textit{$<dark(T(t))>$} is the average value of the dark counts in each pixel at temperature \textit{T} at time \textit{t}, \textit{equ (k)} is the PM0091 gain equalization map for the period in question and whose average value of global efficiency is described by \textit{$<\varepsilon{}_{total}>$}, and  \textit{GF${}_{pixel}$} is the geometric factor of the PM0091 pixels. As described in a previous paragraph, the dark counts are considered constant and a conservative value is adopted (0.7~counts/second per pixel).

\subsection{The relation of the UVscope measurements with sky transparency and global cloudiness}
\label{skytransparence}

To confirm the dependence from the atmosphere conditions of the NSB measurements performed with UVscope we made use of different auxiliary systems available on site, mainly the Sky Quality Meter, SQM, and the All Sky Camera, ASC \cite{Leto 2014}.

The SQM instrument is mounted onto the external support of the ASTRI-Horn secondary mirror; it is sensitive only to visual light and measures, every minute, the magnitude of the sky brightness (mag/arcsec$^{2}$) averaged in a region of 10\ensuremath{^\circ} around the optical axis of the telescope. The ASC instrument, with a FoV of 180\ensuremath{^\circ} and mounted near the telescope, basically provides the global cloud coverage (percentage) every 2~minutes.

%Fig.\ref{nsb,sqm,asc}
\begin{figure}[htb]
\begin{center}
\includegraphics[width=11.5cm]{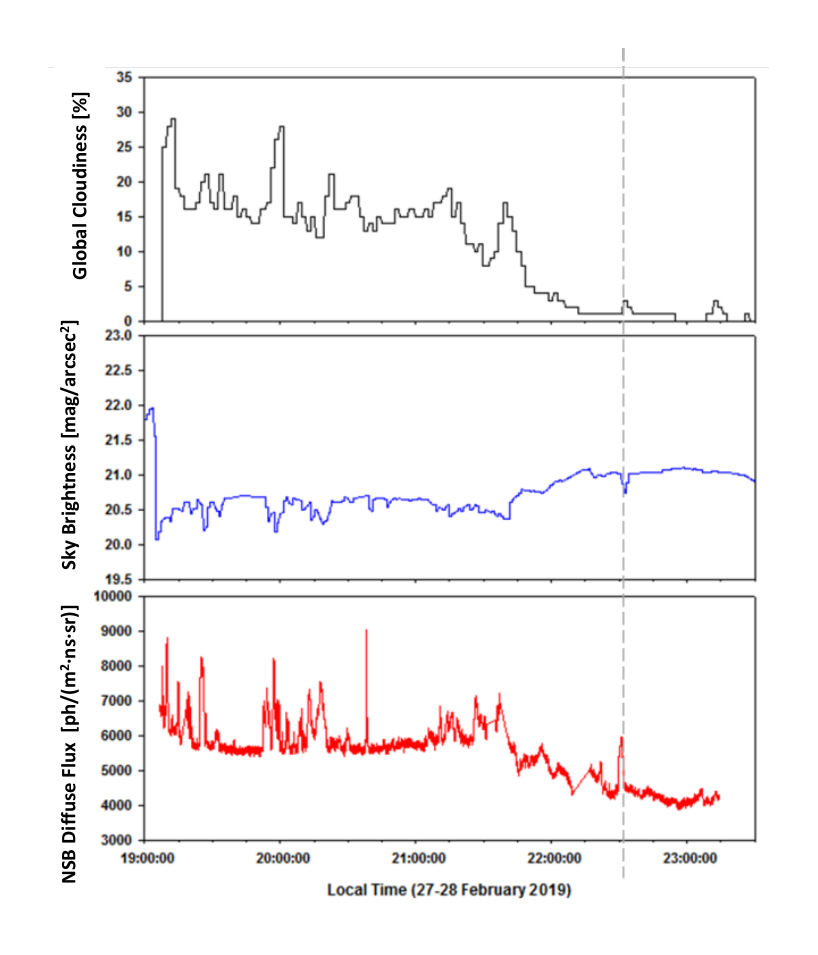}
\caption{Correspondence between diffuse NSB light and sky conditions. From top to bottom, time profile of the global cloudiness (from ASC), the magnitude sky brightness (from SQM) and  diffuse NSB flux (from UVscope) during the night February 27, 2019. The dashed line marks a peak in the diffuse NSB flux that corresponds to a sudden depression of the sky brightness due to, albeit minimal, a peak of the global cloudiness.}
\label{nsb,sqm,asc}
\end{center}
\end{figure}

The diffuse NSB flux profile evaluated during one moonless but cloudy night is shown in Fig.\ref{nsb,sqm,asc}. This night was characterized by strong but decreasing cloudiness and Fig.\ref{nsb,sqm,asc} shows, from top to bottom, the time profile of three measurements: the global cloudiness (ASC), the magnitude sky brightness (SQM) and the diffuse NSB flux (UVscope). As expected, the diffuse NSB flux, evaluated within the dynamic useful mask, increases when the magnitude sky brightness decreases (presence of clouds). Such a behavior is evident along all the time profile; to be noted even the NSB peak, marked by the dashed line, that corresponds to a sudden depression of the sky transparency (SQM) and to a peak, albeit minimal, of the global cloudiness (ASC).
Due to the SQM and ASC functional setting that average their data in time intervals and field of views bigger than those of UVscope, the peak around 20:40:00 local time is strongly evident only in the NSB flux profile evaluated by UVscope. 

%Fig.\ref{nsb,alt,azi}

\section{Observations and analysis}
\label{analysis}

From June 2018 until March 2019, several acquisitions have been performed with UVscope aboard ASTRI-Horn. The summer 2018, with nights always characterized by clear sky, was devoted  to check all UVscope functionalities and the procedure to evaluate the diffuse NSB flux. After a pause due to ASTRI-Horn technical maintenance, the scientific acquisitions restarted in winter 2018. A selection of these acquisitions and their analysis results are described in the following.

During December 2018, the sky was characterized by high variability of several parameters such as cloudiness, relative humidity, presence of volcanic dust (Etna erupting). Volcanic activity was characterized by degassing and continuous explosive activity from the top craters with the formation of ash clouds and lava emissions. These conditions, mainly extreme for the ASTRI-Horn camera, allowed scientific acquisitions only for a portion of the foreseen useful nights. Nevertheless, the period was very fruitful for ASTRI-Horn that obtained its first detection of the Crab Nebula  (R.A.${}_{J2000}$=5h34m31.94s and Dec${}_{J2000}$=+22\ensuremath{^\circ}00'52.2'')  \cite{Lombardi 2020}. 

%Fig.\ref{7-8 December 2018}, Fig.\ref{8-9 December 2018} and Fig.\ref{11-12 December 2018} 

\begin{figure}[htb]
\begin{center}
\includegraphics[width=11.5cm]{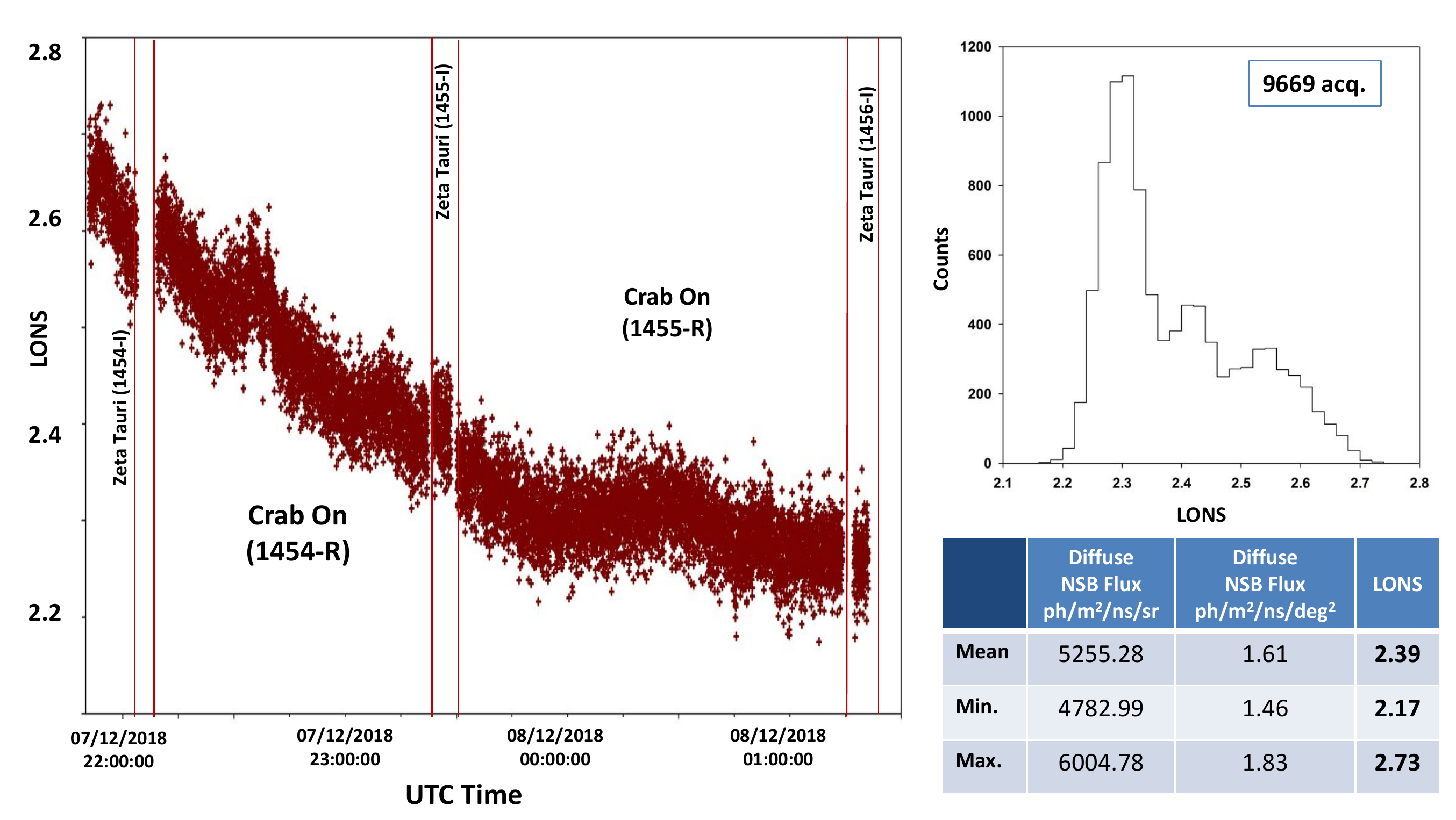}
\caption{Night 7-8 December 2018. Rarely scattered clouds, Etna erupting, column of smoke and ash from the top of the volcano. Statistics of the diffuse NSB flux during Crab and Zeta Tauri observations.}
\label{7-8 December 2018}
\end{center}
\end{figure}

\begin{figure}[htb]
\begin{center}
\includegraphics[width=11.5cm]{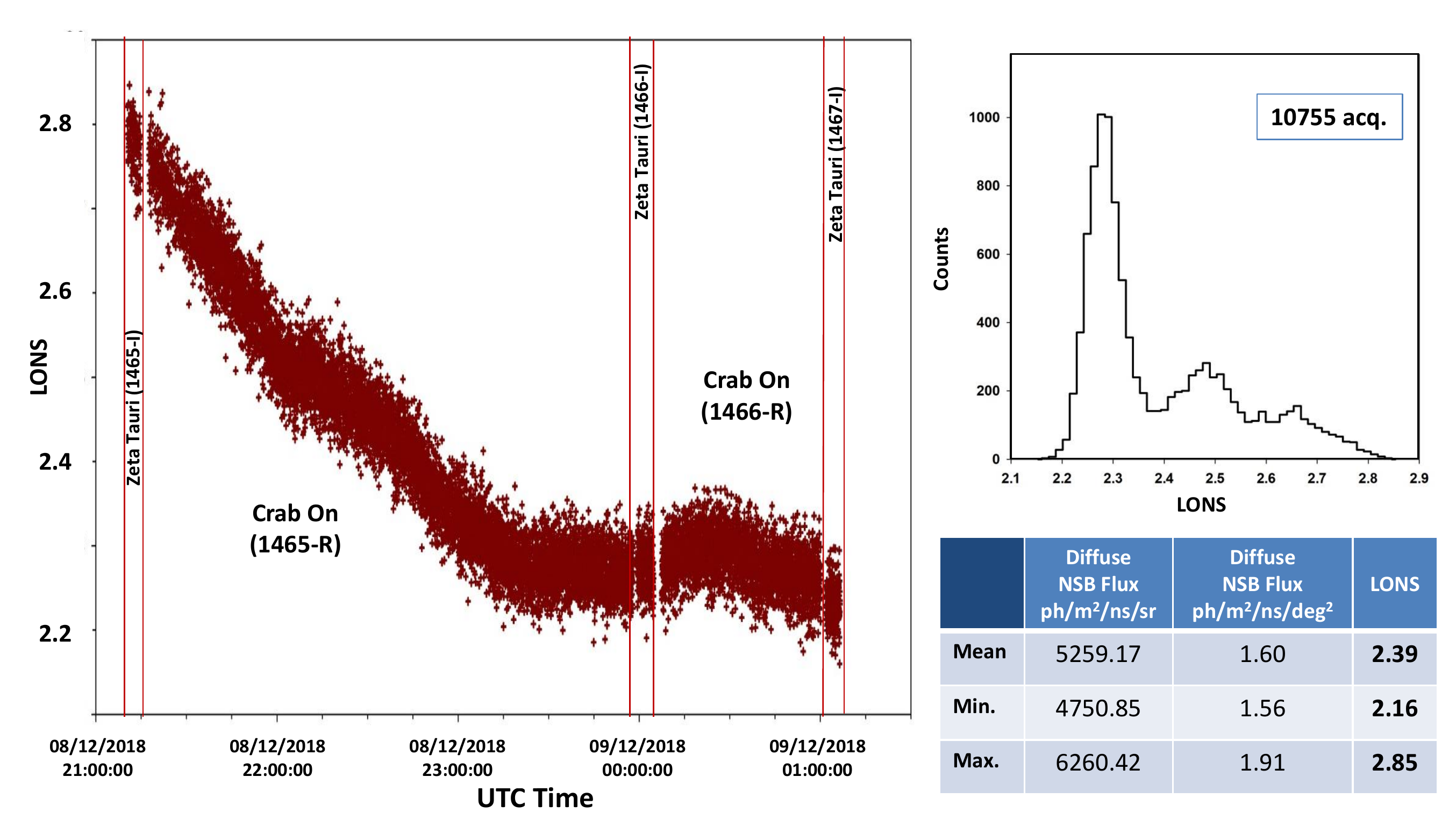}
\caption{Night 8-9 December 2018. Increased cloudiness with respect the previous night; column of smoke and ash from the top of the Etna volcano. Statistics of the diffuse NSB flux during Crab and Zeta Tauri observations.}
\label{8-9 December 2018}
\end{center}
\end{figure}

\begin{figure}[htb]
\begin{center}
\includegraphics[width=11.5cm]{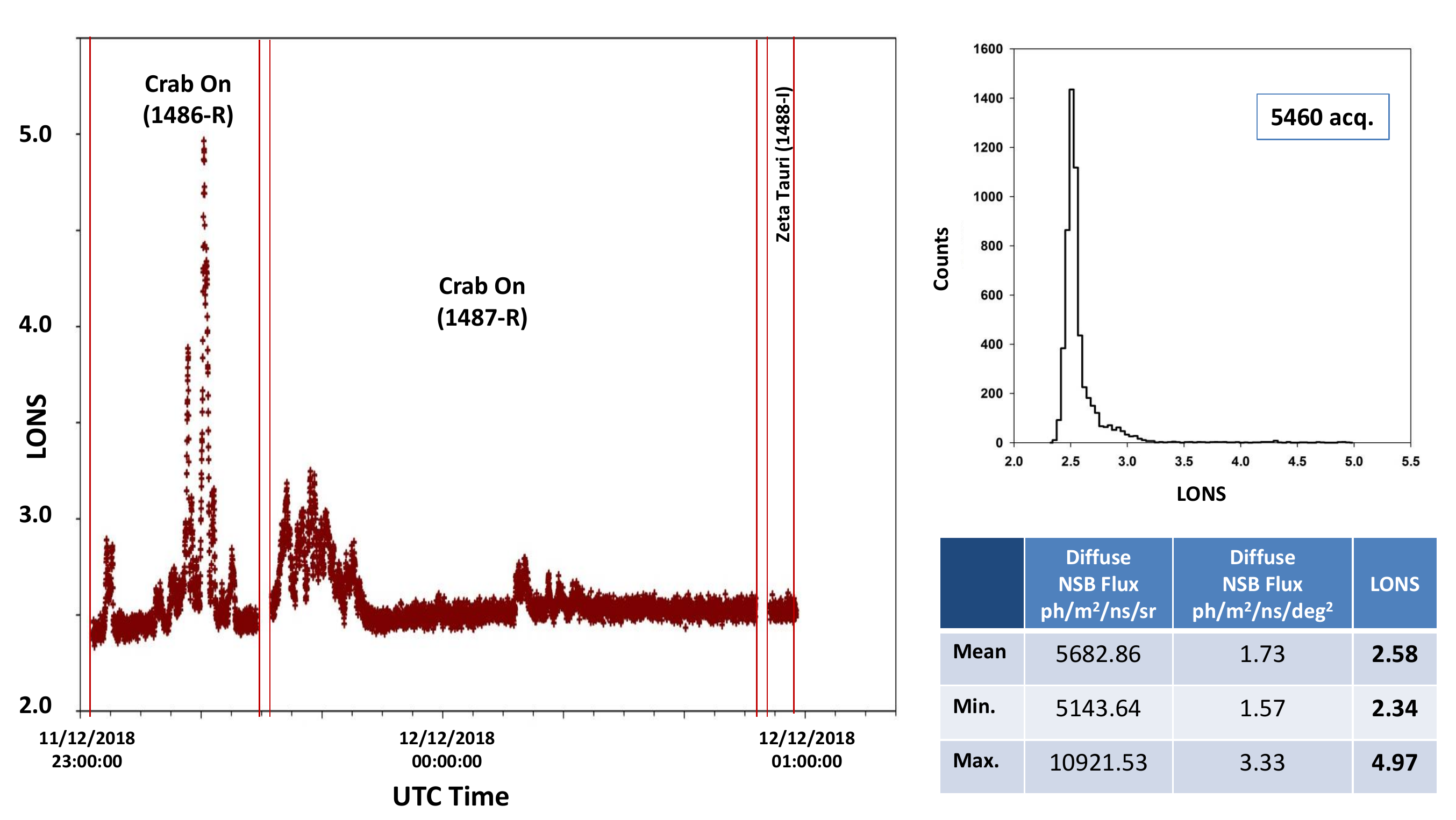}
\caption{Night 11-12 December 2018. Variable and high-level cloudiness, mainly during the first part of the Crab observation; possible presence of volcanic ash. Statistics of the diffuse NSB flux during Crab and Zeta Tauri observations.}
\label{11-12 December 2018}
\end{center}
\end{figure}

Fig.\ref{7-8 December 2018}, \ref{8-9 December 2018} and \ref{11-12 December 2018} show the diffuse NSB flux in the wavelength band of UVscope as obtained by the analysis of the data acquired contemporarily with ASTRI-Horn camera; superimposed is indicated the ASTRI-Horn acquisition run number and the telescope pointing (Zeta Tauri is located at about 1.5\ensuremath{^\circ} far from the Crab Nebula). The figures report the value of diffuse NSB flux in unit of LONS, Light of the Night Sky; as reference, we use the unitary LONS equal to the mean value of diffuse NSB flux detected in La Palma, Canary Islands \cite{Preuss 2002} that, in the wavelength band 300--650~nm, proper of UVscope, corresponds to about
2200~photons/(m$^{2}\cdot$~ns~$\cdot$~sr)~=~0.67~photons/(m$^{2}\cdot$~ns~$\cdot$~deg$^{2}$).

During the first two nights the sky was quite clear with sporadic presence of thin and spare clouds. The diffuse NSB light flux was varying from 2.85 to 2.16~LONS during the Crab Nebula and Zeta Tauri observation (Fig.\ref{7-8 December 2018} and Fig.\ref{8-9 December 2018}). The presence of strongest cloudiness during the third night (Fig.\ref{11-12 December 2018}) increased the diffuse NSB flux till 4.97~LONS.

%Fig.\ref{7-8 December 2018}

%Fig.\ref{8-9 December 2018}

%Fig.\ref{11-12 December 2018}

During the night 8-9 December 2018, under good atmospheric conditions, the ASTRI-Horn telescope followed the Crab Nebula path from 21:00 on the evening of December 8, to 01:00 on December 9, 2018, except a brief observation of the Zeta Tauri star. The Fig.\ref{8-9 December 2018} shows the trend of the diffuse NSB flux as evaluated with UVscope. The first part of Crab's observation (1465-R) is strongly affected by the pointing that, even if more than 50\ensuremath{^\circ} in Altitude, passes over the city of Catania. The observation during the culmination of Crab (Altitude over 70\ensuremath{^\circ}) occurs when the city of Catania has been passed and the level of diffuse NSB decreases and remains constant during the last portion of the acquisition 1465-R to rise moderately during the first part of the subsequent acquisition 1466-R, towards South-West, where there is diffused artificial light due to a small inhabited country.

\section{Conclusions and future plans}
\label{conclusions}

The analysis of the data acquired with UVscope mounted aboard the ASTRI-Horn telescope has confirmed that this compact instrument can give useful information to monitor the night sky background so providing substantial support to test and diagnostic purposes of the novel ASTRI-Horn telescope. It is true that in the ASTRI-Horn system there are other auxiliaries monitoring the sky environment but it is important to highlight that some features makes UVscope an element that is sometimes more selective than the others, as in the case of the SQM instrument, devoted to the night sky brightness monitoring. The SQM is sensitive only to the visual light and averages in a 10\ensuremath{^\circ} field of view around the target and, as a base, records its data at 15 second interval. Under the current configuration, UVscope covers a wavelength region UV extended in a field of the order of 4\ensuremath{^\circ} around the target, a field that corresponds to the central area of approximately 3\ensuremath{\times}3 PDMs of the ASTRI-Horn camera, and records its data at 1 second rate; all these features are really appropriate to support diagnostic purposes of the ASTRI-Horn telescope.

As described in a previous chapter, the firmware of the novel ASTRI-Horn camera continuously performs the statistical analysis of the signal detected by the front-end electronics  and provides the variance of each pixel, which is linearly dependent on the rate of detected photons. The variance technique therefore allows us to indirectly measure the night sky background flux. To validate the variance technique, it is necessary to compare its results with respect to a reference point; in our case, such reference
is provided by the diffuse NSB flux evaluated by the data acquired with the well-calibrated UVscope instrument. This application and its results are the basis of a dedicated paper, currently in preparation.

For the next campaigns, some improvements are expected. To confirm the correct alignment between UVscope and the ASTRI-Horn camera, a collimator of proper length will be temporarily applied in UVscope so to decrease its pixel angular dimension reaching a correspondence one-to-one with the ASTRI-Horn camera pixel.

Moreover, depending on the length of the long collimator, and observing a set of known stars, their flux detected by UVscope can be compared with the ASTRI-Horn results and therefore it should be monitored the photon detection efficiency of the central PDM (or central pixels) of the ASTRI-Horn camera.

A further application of UVscope is devoted to the absolute end-to-end calibration of the ASTRI-Horn camera to be performed with a calibrated illumination system, named Illuminator  \cite{Segreto 2016} externally located at a proper distance from the telescope. A description of the Illuminator is outside the scope of this paper; it is sufficient to recall here few essential points. To measure the telescope end-to-end spectral response, ASTRI-Horn and UVscope will be illuminated simultaneously by a spatially uniform and monochromatic source at several wavelengths provided by the Illuminator. The measurements obtained by the auxiliary UVscope allow to determine the illuminating flux at the ASTRI-Horn telescope aperture and thus completely eliminate the uncertainty due to the atmospheric transmission and propagation; by changing the telescope pointing with respect to the Illuminator, the response of the telescope will be measured with respect to off-axis angles (flat-field). To verify the telescope response to Cherenkov like pulses, a fast pulsed (few nanoseconds) light source will be connected to the Illuminator. Given its small aperture, the auxiliary UVscope, working in single photon counting mode, will provide an accurate calibration of the illuminating flux at the ASTRI-Horn telescope aperture.

Last but not least, it has to be recalled that the ASTRI-Horn telescope is the prototype of nine similar telescopes forming the ASTRI Mini-Array  \cite{Pareschi 2019} that will be installed at the Teide Astronomical Observatory,  in Tenerife, Canary Islands, Spain. In view of the ASTRI Mini-Array, a new auxiliary instrument, named UVSiPM, is in preparation \cite{Sottile 2013} as successor of UVscope, working in SPC mode as its progenitor but equipped with Silicon Photomultiplier sensors, the same model of sensors forming the camera of the ASTRI Mini-Array telescopes. A campaign of comparison between UVSiPM and UVscope is foreseen at the Serra La Nave site and subsequently at the Teide Observatory with both UVSiPM and UVscope configured in portable version.

\begin{acknowledgements}
This work was conducted in the context of the ASTRI Project, supported by the Italian Ministry of Education, University, and Research (MIUR) with funds specifically assigned to the Italian National Institute of Astrophysics (INAF), and by the Italian Ministry of Economic Development (MISE) within the Astronomia Industriale program. This work has gone through internal review by the ASTRI Project Collaboration.
\end{acknowledgements}


\begin{thebibliography}{00}

\bibitem{Giarrusso 2001}
Giarrusso, S., G. Agnetta, B. Biondo, O. Catalano, G. Cusumano, G. Gugliotta, G. La Rosa, M.C. Maccarone, A. Mangano, Fr. Russo, B. Sacco (2001). ``Nocturnal atmospheric UV background measurements in the 300-400 nm wavelenghts band with BaBy 2001, a transmediterranean balloon borne experiment'', In: K.H. Kampert, G. Hainzelmann, C. Spiering (Eds.), `Proc. 27th ICRC Conference', Hamburg, Germany,  Copernicus Gesellschaft,  CD-ROM icc0307\_p, HE, pp. 684-686.

\bibitem{Giarrusso 2003}
Giarrusso, S., G. Gugliotta, G. Agnetta, P. Assis, B. Biondo, O. Catalano, F. Celi, G. Cusumano, G. D'Al\`{i} Staiti, R. Di Raffaele, M.C. Espirito Santo, M. Gabriele, G. La Rosa, M.C. Maccarone, A. Mangano, T. Mineo, M. Pimenta, Fr. Russo, B. Sacco, A. Santangelo, L. Scarsi, B. Tom\'{e} (2003). ``Measurements of the UV Nocturnal Atmospheric Background in the 300-400 nm Wavelengths Band with the Experiment BaBy during a Transmediterranean Balloon Flight''. In: T. Kajita, Y. Asaoka, A. Kawachi, Y. Matsubara and M. Sasaki (Eds.), `Proc. 28th ICRC Conference, Tsukuba, Japan, Universal Academy Press, vol. HE,  p. 849-852.

\bibitem{Catalano 2009}
Catalano, O., G. La Rosa, M.C. Maccarone, A. Segreto, et al. (2009). " UVscope: an instrument for multi-wavelength study of the diffuse Night Sky Background light ". In: `Proc. 31st  ICRC-2009', Lodz, Poland, (CD-ROM, icrc0373). EID: 2-s2.0-84899088153

\bibitem{Catalano 2008}
Catalano, O., M.C. Maccarone, B. Sacco (2008). "Single photon counting approach for imaging atmospheric Cherenkov telescopes", Astroparticle Physics,  vol. 29, n.2, pp. 104-116. DOI: 10.1016/j.astropartphys.2007.11.011. EID: 2-s2.0-39749133703.

\bibitem{Maccarone 2009}
Maccarone, M.C., O. Catalano, G. La Rosa, A. Segreto, et al. (2009). "The UVscope Instrument in the framework of ground-based cosmic ray observatories ". Nuclear Physics B Proc. Suppl, vol.190,  pp. 255-260 (2009). DOI: 10.1016/j.nuclphysbps.2009.03.096. EID: 2-s2.0-65649116522.

\bibitem{Maccarone 2011-nima}
Maccarone, M.C.,  O. Catalano, S. Giarrusso, G. La Rosa, A. Segreto, G. Agnetta, B. Biondo, A. Mangano, Fr. Russo, S. Billotta  (2011). " Performance and applications of the UVscope instrument'', NIM-A, Nuclear Instrum. \& Methods - Section A, vol. 659,  Issue 1, pp. 569-578. DOI:10.1016/j.nima.2011.08.004. EID: 2-s2.0-84860394815.

\bibitem{Maccarone 2011-icrc}
Maccarone, M.C., O. Catalano, S. Giarrusso, G. La Rosa, A. Segreto, G. Agnetta, B. Biondo, A. Mangano, Fr. Russo, S. Billotta (2011). ``Calibration and performance of the UVscope instrument". In: `Proc. 32ndt  ICRC-2011', Bejing, Rep. of China, (CD-ROM, icrc0148). DOI:10.7529/ICRC2011/V03/0148 EID: 2-s2.0-84899535652

\bibitem{Segreto 2011}
Segreto, A., for the Pierre Auger Collaboration (2011). ``Night Sky Background measurements by the Pierre Auger Fluorescence Detectors and comparison with simultaneous data from the UVscope instrument'', In: `Proc. 32ndt  ICRC-2011', Bejing, Rep. of China, Vol. 3 HE1.4, p.129 DOI:10.7529/ICRC2011/V03/0661

\bibitem{Pareschi 2016}
Pareschi, G., for the ASTRI Collaboration and the CTA Consortium, (2016), ``The ASTRI SST-2M prototype and mini-array for the Cherenkov Telescope Array'', Proc. SPIE Astronomical Telescopes + Instrumentation 2016, Paper No. 9906-223.

\bibitem{Scuderi 2018}
Scuderi, S., for the CTA ASTRI Project (2018), ``The ASTRI Program'', Proc. 7th RICAP, 4-7 September 2018, Roma, Italy, EPJ Web of Conferences 209, 01001, https://doi.org/10.1051/epjconf/201920901001.

\bibitem{Horn}
Horn d`Arturo, G. (1953). ``The tessellated mirror'', Journal of the British Astronomical Association 63, 2, pp. 71-74.

\bibitem{Bonoli}
Bonoli, F. (2018), ``Guido Horn d`Arturo and the first multi-mirror telescopes: 1932-1952'', Mem. S.A.It. Vol. 89, 448.

\bibitem{Maccarone 2013}
Maccarone, M.C.,  G. Leto,  P. Bruno,  M. Fiorini,  A. Grillo,  A. Segreto,  L. Stringhetti (2013). ``The Site of the ASTRI SST-2M Telescope Prototype''. In: `Proc. 33rd ICRC-2013', Rio de Janeiro, Brasil, 2-9 July 2013 (CD-ROM id\_0110) (arXiv:1307.5139).

\bibitem{Leto 2014}
Leto, G.,  M.C. Maccarone,  G. Bellassai, P. Bruno,  M. Fiorini,  A. Grillo,  E. Martinetti, G. La Rosa, A. Segreto,  G. Sottile, L. Stringhetti,  for the ASTRI Collaboration (2014). ``The Site of the ASTRI SST-2M Telescope Prototype: Atmospheric Monitoring and Auxiliary Instrumentation''. In: `Proc. AtmoHEAD Workshop 2013', CEA, Saclay, France, 10-12 June 2013 (arXiv:1402.3515v1).

\bibitem{Lombardi 2020}
Lombardi, S., O. Catalano, S. Scuderi, et al. (2020), ``First detection of the Crab Nebula at TeV energies with a Cherenkov telescope in dual-mirror Schwarzschild-Couder configuration: the ASTRI-Horn telescope'', Astronomy \& Astrophysics, Volume 634, id.A22, doi:10.1051/0004-6361/201936791.
\bibitem{Catalano 2018}
Catalano, O., M. Capalbi, C. Gargano, S. Giarrusso, D. Impiombato, G. La Rosa, M.C. Maccarone, T. Mineo, Fr. Russo, P. Sangiorgi, A. Segreto, G. Sottile, B. Biondo, G. Bonanno, S. Garozzo, A. Grillo, D. Marano, G. Romeo, S. Scuderi, R. Canestrari, P. Conconi, E. Giro, G. Pareschi, G. Sironi, V. Conforti, F. Gianotti, R. Gimenes, for the CTA ASTRI Project (2018), ``The ASTRI camera for the Cherenkov Telescope Array'', Proc. SPIE 10702, Ground-based and Airborne Instrumentation for Astronomy VII, 1070237 (6 July 2018), doi: 10.1117/12.2314984.

\bibitem{Pareschi 2019}
Pareschi, G., on behalf of the ASTRI Project, ``The ASTRI Mini-Array at the Teide Observatory'', (2019), presented at the International Workshop on "Future Instrument for the Telescopes at the Observatories de Canarias", Tenerife, Spain, 11-13 Nov. 2019.

\bibitem{Segreto 2019}
Segreto, A., O. Catalano, M. C. Maccarone, T. Mineo, A. La Barbera, for the CTA ASTRI Project (2019), ``Calibration and monitoring of the ASTRI-Horn telescope by using the night-sky background as measured by the photon-statistics  analysis ("variance") of the detector signals'', Proc. 36th ICRC, 24 July -- 1st August 2019, Madison, WI, USA, PoS(ICRC2019)791 Vol.358 (arXiv: 1909.08750).

\bibitem{R7600-M64}
Hamamatsu, "R7600-03-M64 PhotoMultiplier Tube MultiAnode",\\
http://www.sales.hamamatsu.com/

\bibitem{Russo 2006}
Russo, Fr., and G. Agnetta, (2006), ``D.E.L.P.H.IN., Disk Emulator for Laboratory Prototype Hardware Interface'',\\ http://www.ifc.inaf.it/electronic/instruments/delphin/delphinbrochure.pdf

\bibitem{Tanci 2016}
Tanci C. et al., 2016, ``Software design and code generation for the engineering graphical user interface of the ASTRI SST-2M prototype for the Cherenkov Telescope Array'' (2016), Proc. SPIE Astronomical Telescopes + Instrumentation, Paper No. 9913-138

\bibitem{NIST}
"The National Institute of Standards and Technology (NIST)",
http://www.nist.gov

\bibitem{Leinert 1998}
Leinert, Ch., S. Boyer, L.K. Haikaka, M.S. Hanner, A.-Ch. Levasseur-Regourd, I. Mann, K. Mattila, W.T. Reach, W. Schlosser, H.J. Staude, G.N. Toller, J.L. Weiland, J.L. Weinberg, A.N. Witt (1998). "The 1997 reference of diffuse night sky brightness", Astronomy and Astrophysics Supplement Series,  vol. 127, pp. 1-99.

\bibitem{Preuss 2002}
Preuss, S., G. Hermann, W. Hofmann, A. Kohnle (2002), ``Study of the photon flux from the night sky at La Palma and Namibia, in the wavelength region relevant for imaging atmospheric Cherenkov telescope'', {\quad}NIM-A Volume 481, Issues 1--3, 1 April 2002, Pages 229-240.

\bibitem{Segreto 2016}
Segreto, A., M.C. Maccarone, O. Catalano, B. Biondo, C. Gargano, G. La Rosa, Fr. Russo, G. Sottile, M. Fiorini, S. Incorvaia, G. Toso, for the ASTRI Collaboration and the  CTA Consortium (2016),  ``The absolute calibration strategy of the ASTRI SST-2M telescope proposed for the Cherenkov Telescope Array and its external ground-based illumination system'', Proc. SPIE 9906, Ground-based and Airborne Telescopes VI, 99063S (July 27, 2016). Paper No. 9906-142. DOI:10.1117/12.2231922

\bibitem{Sottile 2013}
Sottile, G., Fr. Russo, G. Agnetta, M. Belluso, S. Billotta, B. Biondo, G. Bonanno, O. Catalano, S. Giarrusso, A. Grillo, D. Impiombato, G. La Rosa,  M.C. Maccarone, A. Mangano, D. Marano, T. Mineo, A. Segreto, E. Strazzeri, M.C. Timpanaro (2013). "UVSiPM: a light detector instrument based on a SiPM sensor working in single photon counting ". Nuclear Physics B Proc. Suppl, Volume 239, p. 258-261 (arXiv: 1305.2699). DOI: 10.1016/j.nuclphysbps.2013.05.040.  EID: 2-s2.0-84879868715.

\end{thebibliography}
\end{document}